\documentclass[10pt]{article}
\usepackage{amssymb,amsfonts,amsmath,latexsym}
\usepackage{graphics,graphicx,epsf,epsfig}
\newcommand{\be}{\begin{equation}}
\newcommand{\ee}{\end{equation}}
\newcommand{\bea}{\begin{eqnarray}}
\newcommand{\eea}{\end{eqnarray}}
\newcommand{\ba}{\begin{array}}
\newcommand{\ea}{\end{array}}
\newcommand{\bt}{\begin{tabular}}
\newcommand{\et}{\end{tabular}}

\newcommand{\fr}{\frac}
\newcommand{\ci}{\cite}
\newcommand{\cl}{\centerline}
\newcommand{\bs}{\bigskip}

\newcommand{\en}{\eqno}

\newcommand{\bbib}{\begin{thebibliography}}
\newcommand{\ebib}{\end{thebibliography}}

\begin{document}

\cl{\bf DUALITY AND EFFECTIVE CONDUCTIVITY }

\cl{\bf OF RANDOM TWO-PHASE FLAT SYSTEMS}
\bs

\cl{\bf S.A.Bulgadaev\footnote{e-mail: sabul@dio.ru}}

\bs
\cl{Landau Institute for Theoretical Physics}
\cl{Chernogolovka, Moscow Region, Russia, 142432}

\bs

%\cl{}
\begin{quote}
\footnotesize{
The possible functional forms of the effective
conductivity $\sigma_{e}$ of the randomly inhomogeneous two-phase systems
at arbitrary values of concentrations are discussed.
Two different solutions for effective conductivity are found
using a duality relation, a series expansion
of $\sigma_{e}$ in the inhomogeneity parameter $z$ and some additional
conjectures about the functional form of $\sigma_e$. They differ from the
effective medium approximation, satisfy all
necessary requierements, and reproduce the known formulas for $\sigma_{e}$
in the weakly inhomogeneous
case. This can also signify that $\sigma_{e}$ of the
two-phase randomly inhomogeneous systems may be a nonuniversal function,
depending on some details of the structure of the random inhomogeneities.}
\end{quote}

\vspace{0.5cm}
\cl{PACS: 73.61.-r, 75.70.Ak}
\bs

The electrical transport properties of the disordered systems have
an important practical interest. For this reason they are intensively studied
theoretically as well as experimentally. In this region there is one classical
problem about the effective conductivity $\sigma_{e}$ of inhomogeneous
(randomly or regularly) heterophase system which is a mixture of $N (N \ge 2)$
different phases with different conductivities  $\sigma_i, i =1,2,...,N.$
We confine ourselves here by the simplest case of the two-dimensional heterophase
systems with $N=2.$
Despite of its relative simplicity only a few general exact results have
been obtained so far. Firstly, there is a general expression for
$\sigma_{e}$ in case of weakly inhomogeneous isotropic medium,
when the conductivity fluctuations $\delta \sigma$ are smaller than an average conductivity
$\langle \sigma \rangle$  \ci{1}
$$
\sigma_{e} =
\langle \sigma \rangle \left(1 -
\fr{\langle (\delta \sigma)^2 \rangle}
{D{\langle \sigma \rangle}^2}\right) =
\langle \sigma \rangle \left(1 -
\fr{\langle \sigma^2 \rangle - {\langle \sigma \rangle}^2}
{D{\langle \sigma \rangle}^2}\right),
\en(1)
$$
where $D$ is a dimension of the system.
In our case of two-dimensional two-phase system
$\langle \sigma \rangle = x \sigma_1 + (1-x) \sigma_2, \;
\langle \sigma^2 \rangle - \langle \sigma \rangle^2 = 4x(1-x) {\sigma_-}^2,$
where $x$ is a concentration of the first phase,
$\sigma_- = (\sigma_1 - \sigma_2)/2$,
and (1) takes a form
$$
\sigma_{e} =
\sigma_1\left(1 - (1-x)\fr{2\sigma_-}{\sigma_1} -
\fr{4x(1-x){\sigma_-}^2}{2\sigma_1^2}\right) =
$$
$$
\sigma_+\left(1 + 2(x-1/2)z - 2x(1-x) z^2\right),
\en(2)
$$
where $\sigma_+ = (\sigma_1 + \sigma_2)/2,$ and a new variable
$z = \sigma_-/\sigma_+,$ characterizing an inhomogeneity  of the system,
is introduced.

The further progress in the solution of this problem is connected with
the discovery of a dual transformation, interchanging the phases \ci{2,3}.
This transformation allows to find an
exact formula for $\sigma_{e}$ in the case of systems with equal concentrations
of the phases $x = x_c =1/2$ \ci{3}
$$
\sigma_{e} = \sqrt{\sigma_1 \sigma_2}.
\en(3)
$$
This remarkable  formula is very simple and universal, since it does not depend
on the type of the inhomogeneous structure of the two-phase system.
For systems with unequal phase concentrations a dual transformation gives
a relation between effective conductivities at adjoint concentrations
$x$ and $1-x$ or in terms of a new variable $\epsilon = x-x_c$
($ -1/2 \le \epsilon \le 1/2$) at $\epsilon$ and $-\epsilon$
$$
\sigma_{e}(x, \sigma_1, \sigma_2 )
\sigma_{e} (1-x, \sigma_1, \sigma_2) = \sigma_1 \sigma_2 =
\sigma_{e}(\epsilon, \sigma_1, \sigma_2 )
\sigma_{e} (-\epsilon, \sigma_1, \sigma_2).
\en(4)
$$
Relation (4) means that the product of the effective conductivities
at adjoint concentrations is an invariant. Due to this relation, one can
consider $\sigma_{e}$ only in the regions $x \ge x_c$ ($\epsilon \ge 0$) or
$x \le x_c$ ($\epsilon \le 0$).

However, an explicit formula for the effective conductivity at arbitrary phase
concentrations and z attracts the main interest in this problem.
One such formula was obtained many years ago in the so-called effective
medium (EM) approximation \ci{4}, which turned out to be a good approximation
for random resistor networks not only in the weakly inhomogeneous case \ci{5}.
In this paper, using a duality relation and a series expansion in the
inhomogeneity parameter z, we will find two
explicit approximate expressions for the effective conductivity of two-phase
systems, differing from the EM approximation.
The physical models, corresponding to
them, are introduced in other papers, where their properties are discussed
in detail \ci{6,7}.

Let us start our investigation of the isotropic classical random
two-phase system in the case of arbitrary concentrations with a general
analysis of the possible functional form of the effective conductivity.
Due to the linearity of the defining equations (\ci{1,3}), the effective
conductivity of the random systems must be a homogeneous function of
degree one of $\sigma_i \; (i=1,...,N).$ In the case of $N=2$, it is
convenient to use, instead of $\sigma_i \; (i=1,2),$
the variables $\sigma_+$ and  $z \; (-1 \le z \le 1),$ and, instead of x,
a new variable $\epsilon.$
Then the effective conductivity can be represented in the following form,
symmetrical relative to both phases,
$$
\sigma_{e}(\epsilon, \sigma_+, \sigma_- ) =
\sigma_{+} f(\epsilon, \sigma_-/ \sigma_+) =
\sigma_{+} f(\epsilon, z),
\en(5)
$$
where $\sigma_{e}(\epsilon, \sigma_+, \sigma_- )$ and $f(\epsilon, z)$
must have the next boundary values
$$
\sigma_{e}(1/2, \sigma_+, \sigma_- ) = \sigma_1, \quad
\sigma_{e}(-1/2, \sigma_+, \sigma_- ) = \sigma_2,
$$
$$
f(1/2, z) = 1+z, \quad f(-1/2, z) = 1-z, \quad f(\epsilon, 0) = 1.
\en(5')
$$
The duality relation  in these variables takes the form
$$
f(\epsilon, z) f(-\epsilon, z) = 1-z^2,
\en(6)
$$
from which it follows that at critical concentration $\epsilon = 0$
$$
f(0, z) = \sqrt{1-z^2}.
\en(3')
$$
Strictly speaking, the form of a duality relation (6) is also a consequence
of another exact relation for the effective conductivity,
taking place at arbitrary concentrations for systems with
the similar random structures of both phases of the system,
$$
\sigma_{e}(\epsilon, \sigma_1, \sigma_2 ) =
\sigma_{e} (-\epsilon, \sigma_2, \sigma_1).
\en(7)
$$
This means that the effective conductivity of the random two-phase system
must be invariant under substitution of these phases
($\sigma_1 \longleftrightarrow \sigma_2$) with the corresponding
change of their concentrations $x \longleftrightarrow 1-x$ (or $\epsilon
\to -\epsilon$). In the new variables this means that
$$
f(\epsilon, z) = f(-\epsilon, -z),\quad f(-\epsilon, z) = f(\epsilon, -z).
\en(8)
$$
For this reason, the duality relation can also be written  in the form
$$
f(\epsilon, z) f(\epsilon, - z) = 1-z^2.
\en(9)
$$
It follows from (8) that the even ($f_s$) and odd ($f_a$) parts of
$f(\epsilon,z)$ relative
to $\epsilon$  coincide with the even ($f^s$) and odd ($f^a$) parts of
$f(\epsilon,z)$ relative to $z.$
Consequently, $f(\epsilon,z)$ has  the functional form
$$
f(\epsilon,z) = f(\epsilon z, \epsilon^2, z^2).
\en(10)
$$
It follows from (10) that: (1) $f(0,z)$ is an even function of $z$ (i.e.
symmetric in $\sigma_{1,2}$), (2) an expansion of $f(\epsilon,z)$ near the
point $\epsilon = z = 0$
does not contain terms linear in $\epsilon$ and $z$ separately.
Analogously, the odd part $f_a$ can be represented in the form
$$
f_a(\epsilon,z) = 2\epsilon z \Phi(\epsilon, z),
\en(11)
$$
where $\Phi$ is an even function of $\epsilon$ and $z$ (the coefficient $2$
in front of $\epsilon z$ is chosen for further convenience).

At first sight, the duality relation (11) alone is not enough
for the complete determination of $f$ in the general case. It only connects
$f$ at adjoint concentrations or $f_a$ and $f_s$
$$
f_s^2 - f_a^2 = 1-z^2.
\en(12)
$$
This means that $f_a$ and $f_s$ considered at fixed $z$ as the functions
of $\epsilon$ satisfy the hyperbolic relation with a constant depending on
$z.$ The relation (12) allows one to express $f(\epsilon,z)$ through its
even or odd parts
$$
f(\epsilon,z) =
f_a + \sqrt{f_a^2 + 1-z^2} =
f_s \pm \sqrt{f_s^2 - 1 + z^2}.
\en(13)
$$
For this reason, it is enough to know only one of these two parts.
Usually, one prefers to choose an antisymmetric part as a more simple one.
It follows from (2) that, in the weakly inhomogeneous case, the odd part
coincides with the odd part of $\langle \sigma \rangle$ and has
the simplest form (compatible  with (11))
$$
f_a(\epsilon,z) = 2 \epsilon z.
\en(14)
$$
As is well known,  the effective conductivity in the EM approximation
can be obtained by substitution of (14) into (13):
$$
\sigma_{e}(\epsilon,z) = \sigma_+ \left[2 \epsilon z +
\sqrt{(2 \epsilon z)^2 + 1 - z^2}\right].
\en(15)
$$
We will call this expression, continued on arbitrary concentrations
$x = \epsilon + 1/2$ and inhomogeneties $z,$ the EM approximation for
$\sigma_e$.

However, systems with a dual symmetry usually have some additional hidden
properties, permitting one to obtain more information about function under
question. Moreover, in some cases these properties can help to solve problem
exactly (see, for example, \ci{8}). Having this in mind, we will try to
investigate the duality relation in more detail.
For every fixed $z \ne 1$ (it is enough to consider only the region
$0 \le z \le 1$), a function $f$ must be a monotonous function of $\epsilon.$
Since the homogeneous limit $z \to 0$ is a regular point of $f,$ it will be very
useful to expand $f$ in a series in $z$
$$
f(\epsilon, z) = \sum_0^{\infty} f_k(\epsilon) z^k/k!,
\en(16)
$$
where due to the boundary conditions (5')
$$
f_0 = 1, \quad f_1(\epsilon) = 2 \epsilon.
\en(17)
$$
It is worth noting here that the expansion (16) differs from the weak-disorder
expansion of $\sigma_e$ in series on the averaged powers of the conductivity
fluctuations $\delta \sigma/\langle \sigma \rangle$ (see, for example \ci{9}).
Expansion (16) is simpler, since it deals with variables z and $\epsilon$
separately, while the expansion in powers of
$\delta \sigma/\langle \sigma \rangle$ is an expansion on the rather
complicated functions of z and $\epsilon.$  Of course, both expansions are
connected, but expansion (16) is more convenient for our analysis of possible
functional forms.

Substituting the expansion (16) into (6) one obtains
the following results:

(1) in the second order on $z$ it reproduces a universal formula (2),
thus the latter can be considered as a consequence of the duality relation;

(2) in higher orders there are recurrent relations between $f_{2k}$ and
$f_{2k-1},$ corresponding to connection (12);

(3) $f_{2k+1}(\epsilon)$ are odd polynomials in $\epsilon$ of degree
$2k+1$ and
$f_{2k}(\epsilon)$ are even polynomials in $\epsilon$ of degree $2k$
in agreement with (10).

Taking into account boundary conditions (5') and an exact value
(3'), one can show that the coefficients $f_{k}$ must have the next form
$$
f_{2k+1}(\epsilon) = \epsilon (1-4\epsilon^2) g_{2k-2}(\epsilon),
\quad k \ge 1,
$$
$$
f_{2k}(\epsilon) = (1-4\epsilon^2) h_{2k-2}(\epsilon),
\quad k \ge 1,
\en(18)
$$
where $g_{2k-2}$ and $h_{2k-2}$ are some even polynomials of the corresponding
degree and free terms of $h_{2k-2}$ coincide with the coefficients in the
expansion of (3')
$$
\sqrt{1-z^2} = 1-z^2/2 -z^4/8 -z^6/16 -z^8/128 -z^{10}/256 + ...
\en(19)
$$
It follows from (18) that $f_3$ is completely determined up to overall
factor number $g_0.$ Since $f_4$ is determined through lower
$f_k \quad (k=1,2,3)$
$$
f_4 = 4f_1 f_3 - 3f_2^2 = (1-4\epsilon^2)[(8g_0 + 12)\epsilon^2 - 3],
\en(20)
$$
it is also determined by the coefficient $g_0.$ Expansion (16)
in the EM approximation has a very simple form, since all
$g_{2k-2}=0 \; (k \ge 1)$ and $f_{2k}(\epsilon) \sim (1-4\epsilon^2)^k$.
Thus, we see that, in general case, the arbitrariness of $f$ is strongly reduced
by boundary conditions and by exact value (3) and that the third and fourth
orders are determined only up to one constant. One can see from the EM
approximation  that
any additional information about function $f$ can determine this constant
or even the whole function.
For this reason one needs to know what kind of functions can
satisfy the duality relation (6) except general functions from (12),(13).
In order to answer this question
it is convenient in the case $z \ne 1$ to pass from $f$ to
$\tilde f = f/\sqrt{1-z^2}.$ Then
$$
\tilde f(\epsilon, z) \tilde f(-\epsilon, z) = 1 =
\tilde f(\epsilon, z) \tilde f(\epsilon, -z).
\en(6')
$$
The duality relation gives some constraints
on the possible functional form of $\tilde f(\epsilon,z).$
For example, assuming a functional form (10), one can write out
the next simple expression:
$$
\tilde f(\epsilon,z) = \exp(\epsilon z \phi(\epsilon,z)),
\en(21)
$$
where $\phi(\epsilon, z)$ is some even function of its arguments.
Another possible form of $\tilde f$ is
$$
\tilde f(\epsilon,z) = B(\epsilon z)/ B(-\epsilon,z)).
$$
It is easy to see that they automatically satisfy eq.(6').

Let us now consider two simple ansatzes for a function $\phi.$ In case (a) we
suppose that $\phi(\epsilon, z)$ depends
only on $z.$ This means an exponential dependence on concentration, which
sometimes takes place in disordered systems \ci{10}. In case (b) we will
suppose that $\phi(\epsilon, z)$ depends only on the combination $\epsilon z.$
This can signify, for example, that $f$ depend only on mean conductivity
$\langle \sigma \rangle$ and/or on mean resistivity
$\langle \sigma^{-1} \rangle,$ since
$\langle \sigma^{\pm} \rangle \sim (1\pm 2\epsilon z).$
Expanding the corresponding functions $\tilde f$ in series, one can check
after some algebra that it is now  possible to determine all polynomial
coefficients unambiguously!
For example,  one finds for $f_a$ in the 3-d and 5-th orders
$$
g_0 = -1, \quad g_2 = -(11+4\epsilon^2), \; case (a),
$$
$$
g_0 = -3, \quad g_2 = -15(1+12\epsilon^2), \; case (b).
$$
Another way to see this is to apply boundary conditions directly to the
function (21). In the case (a) one obtains
$$
\phi(z) = 1/z \ln \fr{1+z}{1-z}, \quad
\tilde f (\epsilon,z) =  \left(\fr{1+z}{1-z}\right)^{\epsilon}.
\en(22)
$$
It is interesting to note that, in terms of concentration $x$ and partial
conductivities $\sigma_i,$ one obtains in this case
$$
\sigma_e = \sigma_1^x \sigma_2^{1-x}.
\en(22')
$$
This corresponds to the self-averaging of $\ln \sigma:$
$$
\sigma_e = \exp \langle \sigma \rangle, \quad
\langle \sigma \rangle = x \ln \sigma_1 + (1-x) \ln \sigma_2,
$$
noted first by Dykhne for the case of equal phase concentrations \ci{3}
and established later in the theory of weak localization \ci{11}.

In case (b), when $\phi$ depends only on the combination $\epsilon z$,
one finds
$$
\phi(\epsilon z) = \fr{1}{2\epsilon z} \ln \fr{1 + 2\epsilon z}{1-2\epsilon z},
\quad \tilde f(\epsilon,z) =
\left(\fr{1+ 2\epsilon z}{1-2\epsilon z}\right)^{1/2}.
\en(23)
$$
In terms of $x$ and $\sigma_i$ it has the next simple form
$$
\sigma_e = \sqrt{\langle \sigma \rangle/ \langle \sigma^{-1} \rangle}.
\en(23')
$$
Series expansions of (22) and (23) coincide exactly with the corresponding
expansions mentioned above. They differ from the EM approximation
already in the third order.

For a general form of $\phi(\epsilon,z),$
admitting a double series expansion in $z^2$ and $\epsilon^2,$
$$
\phi(\epsilon,z) = \sum_0^{\infty} \phi_k(\epsilon) z^{2k}/k!, \quad
\phi_k(\epsilon) = \sum_0^{\infty} \phi_{kl} \epsilon^{2l}/l!,
$$
one can show that now  $f_3$ and $f_4$ again contain one free parameter
$\phi_{10}:$ $g_0 = 6(\phi_{10} -1).$ Consequently,  one needs
additional information or a more complicated ansatz for a determination of
$\phi$ in the general case.
This will be considered in another paper.

Thus we have found two explicit functions (22) and (23), which satisfy
all required properties. In particular,  they
reproduce equation (2) in the weakly inhomogeneous limit $z \ll 1.$
These functions  can be considered as  regular solutions of the duality
relation, since they are represented by convergent series in $z$  for
$0 \le z \le 1$ except the small region $z \to 1, \epsilon \to 1/2.$

The systems, having the effective conductivity just of two  forms found
above and their properties are considered in the other paper \ci{6}
(see also \ci{7}). We give here only their brief description.

The first model represents randomly inhomogeneous systems with compact
inclusions of the second phase with finite maximal scale $l_m$ of
inhomogeneities. This scale can depend on concentration of the second phase
$l_m (1-x)$ (one can consider only the case $1-x \le 1/2).$ The stable
effective conductivity $\sigma_e(x,\{\sigma\})$
(here $\{\sigma\} = (\sigma_1,\sigma_2)$), depending only $x$ and not
depending on the scale on which the averaging is performed, can be obtained
only after averaging over scales $l > l_m(x).$ This $\sigma_e(x,\{\sigma\})$
as a function of $x$ must satisfy the next functional equation, generalizing
duality relation (4)
$$
\sigma_e(x',\{\sigma\}) \sigma_e(x",\{\sigma\} =  \sigma_e^2(x,\{\sigma\},
\en(24)
$$
where $x=(x'+x")/2.$ The solution of equation (24), satisfying boundary
conditions (5'), coincides with (22) and corresponds to the
finite maximal scale averaging approximation (FMSA) \ci{6,7}.

The second model of a random inhomogeneous systems has a hierarchical,
two-level structure. On the first level, it consists of squares with random
phase layers with a mean conductivity $\langle \sigma \rangle$ if the direction
of layers is parallel to the applied electrical field ${\bf E}$ or with a
conductivity $\langle \sigma^{-1} \rangle^{-1}$ if this direction is
perpendicular to ${\bf E}.$ On the second level, these squares form a random
parquet (or a lattice), which contains with equal probabilities $(p=1/2)$
squares with both orientations. Then, using universal formula (3), one can
write the next approximate expression for $\sigma_e$
$$
\sigma_e(x,\{\sigma\}) =
\sqrt{\langle \sigma \rangle \langle \sigma^{-1} \rangle^{-1}},
$$
which coincides with (23).

For a comparison of the different expressions for effective conductivity
(eqs. (15),(22) and (23)), we have constructed three
plots of the corresponding functions $f(\epsilon,z)$ at $z=0.8, 0.95, 0.999$
(Fig.1) (their full 3D plots are represented in \ci{7}).
\begin{figure}[t]
\centerline{
\begin{tabular}{cc}
{\input epsf \epsfxsize=5.5cm \epsfbox{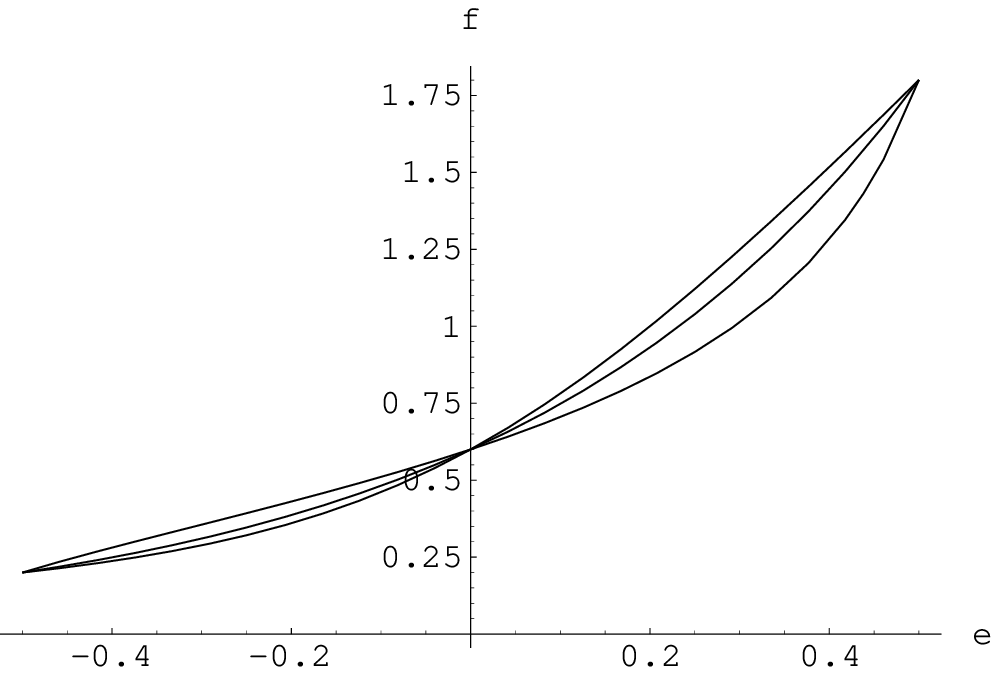}} &
{\input epsf \epsfxsize=5.5cm \epsfbox{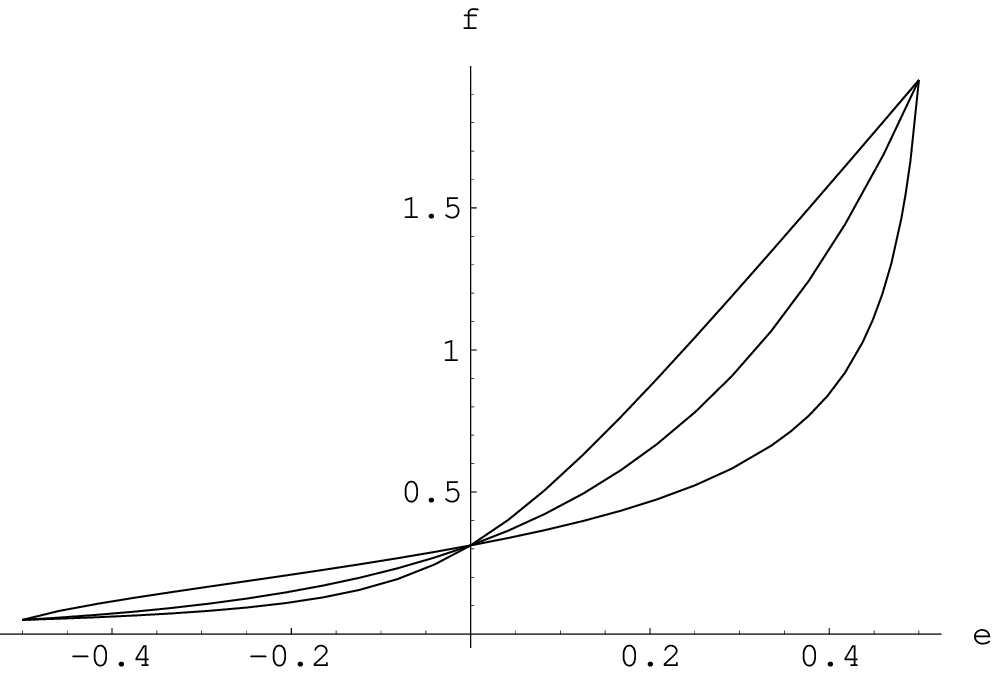}}\\
{} & {}\\
a & b\\
{} & {}\\
\end{tabular}}
\centerline{
\input epsf \epsfxsize=6cm \epsfbox{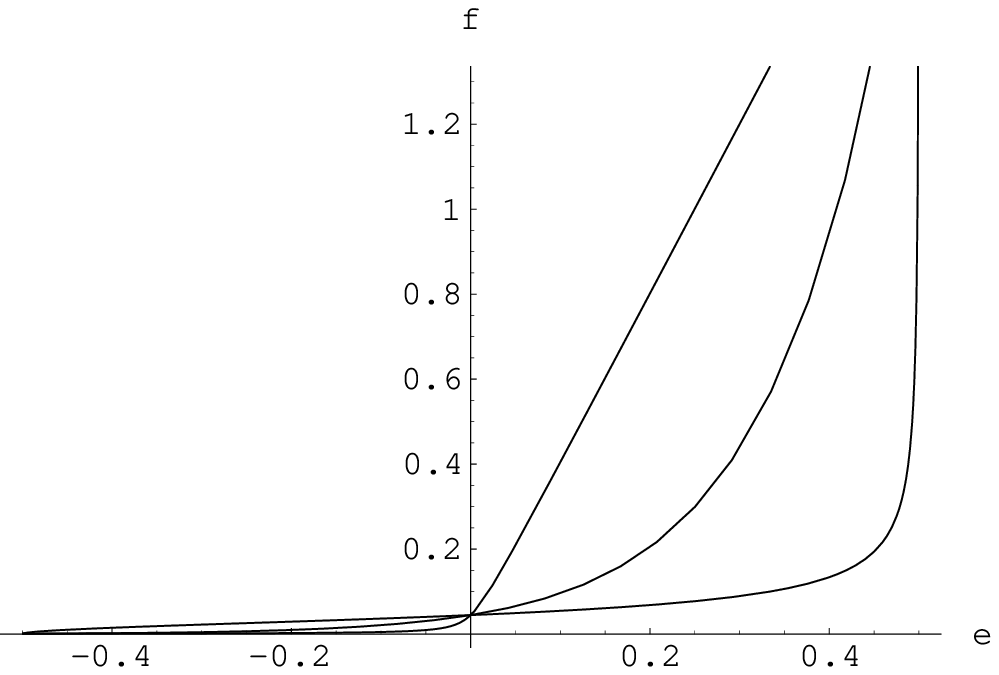}}
\vspace{0.3cm}
\centerline{c}
\vspace{0.3cm}
{\small Fig.1. Plots of various expressions for $f(\epsilon,z)$ at:
(a) $z = 0.8,$ (b) $z=0.95,$ (c) $z=0.999$.}
\end{figure}
The lower branch in the region $\epsilon >0$ corresponds to $f$ from (23),
the upper branch  to the  EM
approximation, and the middle branch  to $f$ from (22).
It appears that all three formulas for $f(\epsilon,z)$,
despite of their various functional forms,
differ from each other very weakly  for
$z \lesssim 0,5$ due to very restrictive boundary conditions (5') and
the exact Keller-Dykhne value. This range of $z$ corresponds approximately
to the ratio $\sigma_2/\sigma_1 \sim 1/3.$
For the smaller ratios, the differences between these functions become
distinguishable (for $\epsilon > 0$), growing significantly only for ratios
$\sigma_2/\sigma_1 \lesssim 10^{-1}.$

One can see from formulas (22),(23) that, in both cases,
one gets  $\sigma_{e} \to 0$ in the limit  $\sigma_2 \to 0$,
except the small region near $x=1$ and $z=1.$
This means that these formulas are not valid in the percolation limit
$\sigma_2 \to 0$  ($z\to 1$) for $\epsilon > 0$ \ci{10,12}.
One can show that such behaviour is a consequence of the  assumptions
made about the form of the function $\phi$ and/or of the structure of the
corresponding models \ci{7}.

This can be connected also with a possible divergence
of the series (16) in general case, when $z \to 1,$ due to a singular
behaviour of $\sigma_{e}$ in the percolation problem \ci{9,10}.

For this reason the formulas (22,23) cannot be
applicable for the description of $\sigma_{eff}$ in the limit
$z\to 1 \, (\epsilon > 0)$ and
the corresponding percolation problem. It follows also from the constructed
plots that EM approximation overestimates $\sigma_{e}$ \ci{10,12}, and both
the other formulas underestimate it in the region $z \to 1, \epsilon > 0.$
We hope to investigate this limit in detail later.

Thus, we have discussed  possible functional forms of the effective
conductivity of random two-phase systems at arbitrary values of concentrations.
It was shown that the duality relation and some additional assumptions
about possible functional form of $f(\epsilon,z)$ can give
its explicit expressions, differing from EM approximation.
They automatically satisfy the duality relation and
reproduce all known formulas for $f$ in the weakly inhomogeneous limit
$z \ll 1.$

Though the used additional assumptions are the approximate ones the obtained
 results (and especially an existence of the corresponding models \ci{6,7})
can be interpreted also  as if $\sigma_{e}$ of the
two-phase randomly inhomogeneous systems were a nonuniversal function,
depending on some details of the structure of the random inhomogeneities.
An analogous conclusion was made earlier for three-phase {\it regular}
systems in  \ci{13}, where a possibility to find a generalization of the
Keller -- Dykhne formula (3) for the case $N = 3$ was studied numerically.

\bs

Acknowledgements

\bs

The author thanks the referees for useful remarks.
This work was supported by the RFBR grants 2044.2003.2 and 02-02-16403.

\end{document}